Word Count: 1832

# The Neural-Wave Quick Escape Manual 2036: A Field Guide to Adversarial Living in the Era of "Empathic" AIoT

Boyuan Gu* [1 2]

Shuaiqi Cheng [1]

**Insights:**

· **"Empathic" smart care systems can prioritize algorithmic compliance over actual human autonomy.**

· **True privacy against camera-free mmWave sensing demands "adversarial literacy" to strategically obfuscate data.**

· **Design fiction exposes how reintroducing friction into seamless computing reclaims domestic dignity.**

---

*correspondence: guboyuan79@gmail.com. [1] Glasgow College, University of Electronic Science and Technology of China (UESTC). [2] Shenzhen International Graduate School (SIGS), Tsinghua University.

## 1 SCENARIO WORLD: 2036 CARE COMPLIANCE

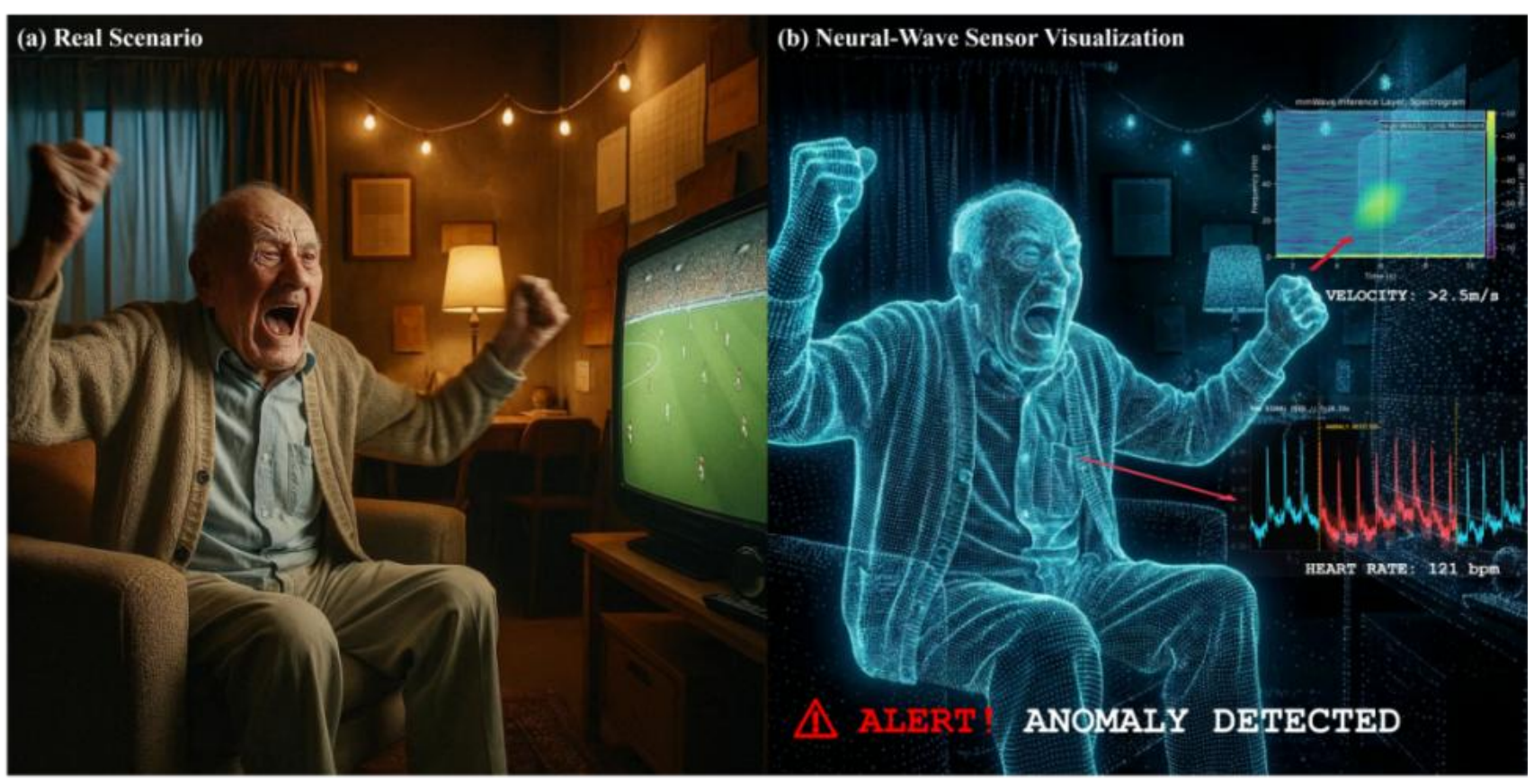

Figure 1. The Ontological Gap in Neural Wave Surveillance. Arthur's joy is algorithmically misdiagnosed as "Pre-Stroke Agitation".

It is 20:40 in a Smart-Care apartment, and Arthur, an 81-year-old retired structural engineer, is watching a live UEFA European Championship match. The game goes into stoppage time, the referee awards a penalty, and Arthur leaps up, shouting. His heart rate rises—not as a pathology, but as fandom, the last reason for an older man to feel urgently alive.

To the Neural-Wave system embedded in his ceiling, however, this joy is illegible. The 60GHz millimeter-wave (mmWave) radar array sees only a volumetric disturbance: a spiking Micro-Doppler signature and a sharp drop in Heart Rate Variability (HRV) that aligns perfectly with its pre-trained model for "Pre-Stroke Agitation". To be fair, these systems are engineered with life-saving intentions—they can successfully detect silent strokes and prevent fatal falls when human caregivers are absent. However, this safety comes at a steep price.

A soft chime wakes the wall interface with polite inevitability: "High stress detected. Please sit, place both feet on the floor and inhale for 4 seconds…". Arthur laughs and swears at the ceiling, shouting that he is fine. But for "privacy," the microphone is disabled; the system can only accept measurable, biometric compliance. Within seconds, the apartment enters Safety Lockdown Mode. The lights dim, and the door locks. The system that claims to protect him has successfully prevented him from walking three meters away from his own television.

Defeated, Arthur sits down to perform calmness for an invisible evaluator. He slows his breathing, stills his trembling hands, and watches the progress ring advance when his body "agrees" with the algorithm. By the time the door unlocks, the match is over. The system logs the incident not as a false alarm, but as a "Successful Intervention: Subject Stabilized".

By 2036, this is what "aging in place" looks like. Driven by chronic shortages of care labor, domestic care has been recast as an auditable public obligation where human lives are prioritized for administrative metrics over interpersonal support [1, 2]. The home is reclassified as a sensing endpoint expected to generate defensible traces of well-being. While early camera-based monitoring triggered obvious resistance because privacy loss was highly visible, the industry simply pivoted to "camera-free" spectral surveillance. Marketed as unobtrusive and privacy-preserving, mmWave technology captures involuntary micro-motions and vital signs through household obstructions, outputting abstract risk scores rather than verifiable footage.

Yet, as Arthur's lockdown demonstrates, "camera-free" never means governance-free. The conflict has merely shifted from "I am being watched" to "I am being inferred" [2]. Under this regime of infrastructural inference, the traditional privacy opt-out has been replaced by a chilling new default: "This feature is unavailable for your safety". To unpack the lack of resident-facing tools to contest this invisible governance, we employ speculative design and fiction [3] to pose

"what if" questions about seamless care, introducing the Neural-Wave Quick Escape Manual—an artifact that proceduralizes the struggle for autonomy in an algorithmically managed home.

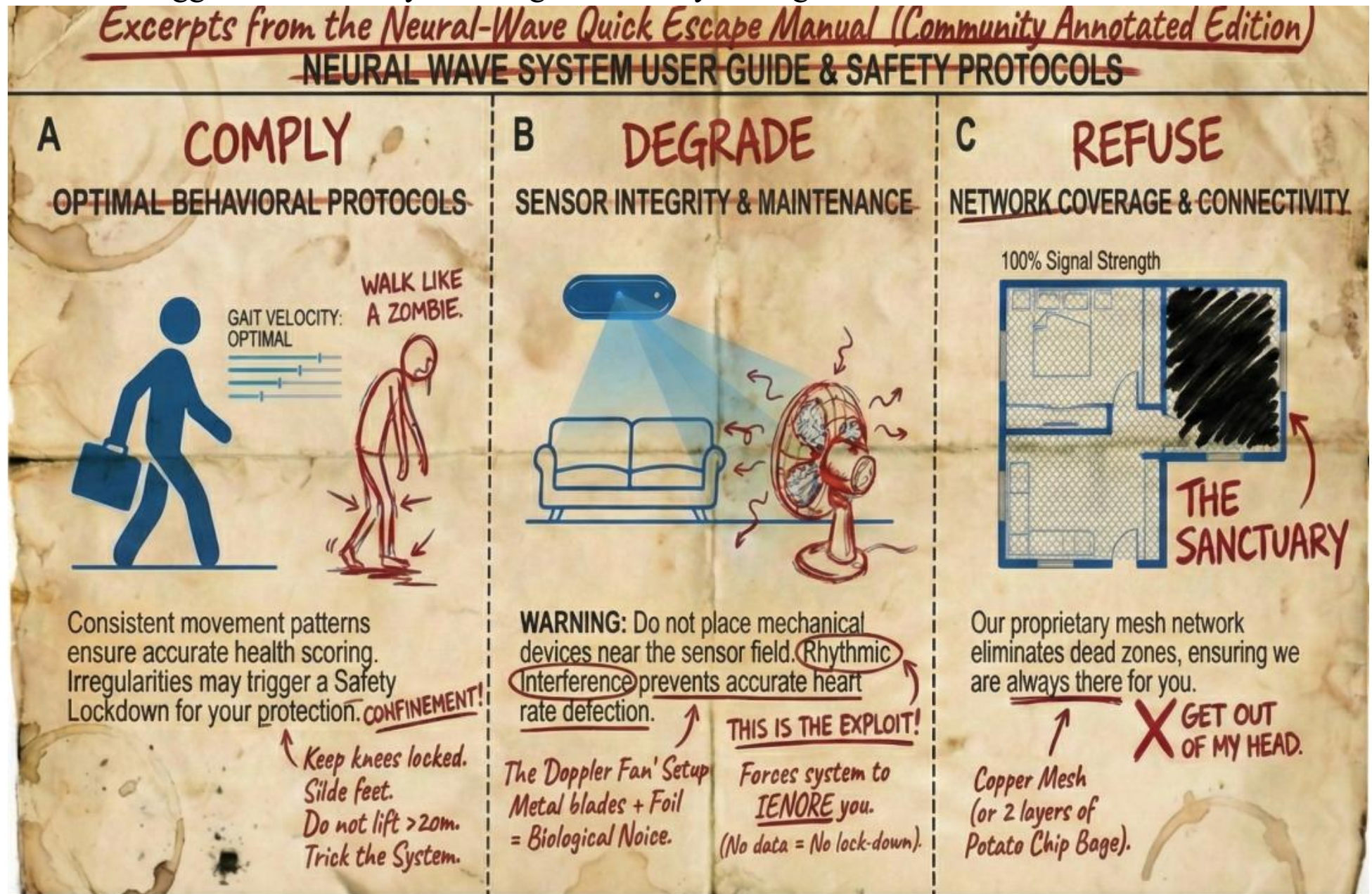

Figure 2. Tactical Resistance Strategies. Excerpt from the fictional Neural-Wave Escape Manual.

## 2 THE ARTIFACTS: WEAPONS OF THE WEAK

Enter the Neural-Wave Quick Escape Manual—an everyday form of non-compliance and sabotage, functioning as a "weapon of the weak" [4] disguised as official paperwork. It is a set of counterfeit pages that treats "privacy" not as a political guarantee, but as an operational problem to be procedurally managed. In the Manual's clinical, detached voice, the resident is reduced to a biological unit whose legitimacy is continuously negotiated through machine-legible traces. There is no outrage, no protest; there is only "standard procedure."

The manual compresses survival into three escalating tactical moves: Comply teaches the body to output an acceptable baseline of calm; Degrade crowds the system with a cleaner proxy signal than the self; and Refuse builds a small, stubborn dead zone where inference simply cannot reach. Read together, these pages do not merely celebrate evasion. They stage a satirical inversion in which "care" becomes infrastructural policing, and dignity is recovered only by learning to perform, to substitute, or to disappear completely.

### Stabilization Protocol — A Tutorial in Biometric Obedience

This protocol presents compliance not as consent but as training: a soft, municipal ritual for producing the "correct" biosignal. Residents are instructed to adopt a rigid stance and execute a vagal-compression routine—bracing the chest to physically constrain respiratory excursion. As the wall interface's 'Stabilization Progress' ring advances only when the body converges with the model's prior, autonomy is recoded as a performance audit. The user earns back ordinary permissions (e.g., unlocking the door) solely by demonstrating signal-level obedience. Thus, the safest way to live under "empathic" sensing is to become legible on demand—even if calmness is merely an output rather than an experience.

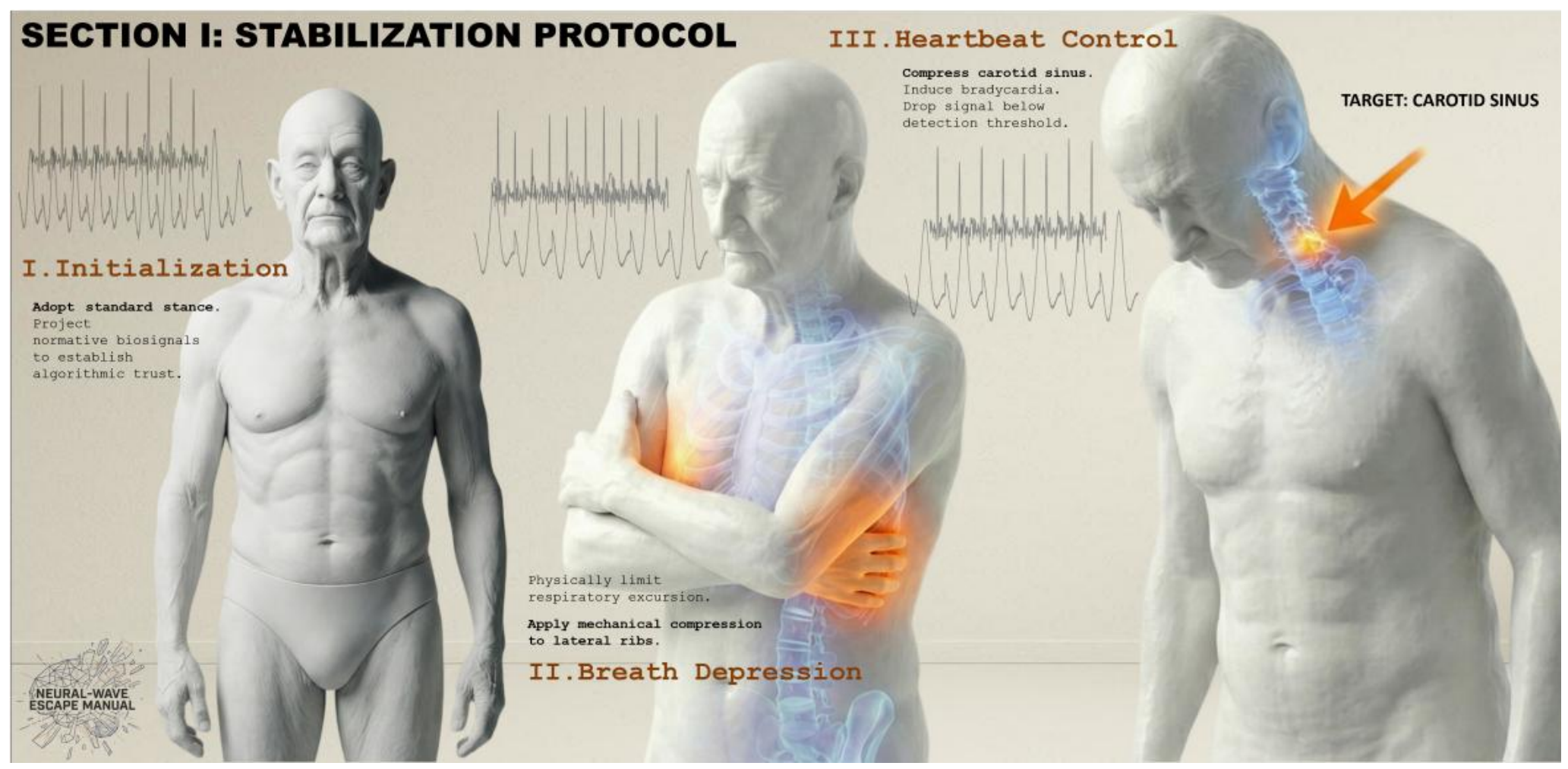

Figure 3. Tactic 01: Bio-Obedience. The user performs the "Statue Protocol," physically suppressing vital signs to appease the algorithm. Here, calmness is not a feeling, but a labor of submission.

**Data Poisoning — Making Normality Louder Than Reality**

While Comply demands the body perform, Degrade outsources this labor through strategic obfuscation [5], protecting privacy not by hiding, but by polluting the data stream with noise. This strategy exploits the radar's reliance on Micro-Doppler signatures via a sensor-spoofing mimicry device. By precisely matching the fundamental frequencies of a resting human heartbeat and respiratory rate (e.g., 1.2 Hz and 0.3 Hz), a highly reflective mechanical oscillator generates a stable,high-amplitude signal. The sensing algorithm, programmed to lock onto the strongest signal, prioritizes this "Phantom" over the chaotic human resident. Hiding in the device's electromagnetic shadow, the user becomes statistically invisible. In this configuration, the smart home's "perfect citizen" is a vibrating piston; the human is rendered safe only by becoming secondary to a machine that performs "normality" louder than reality.

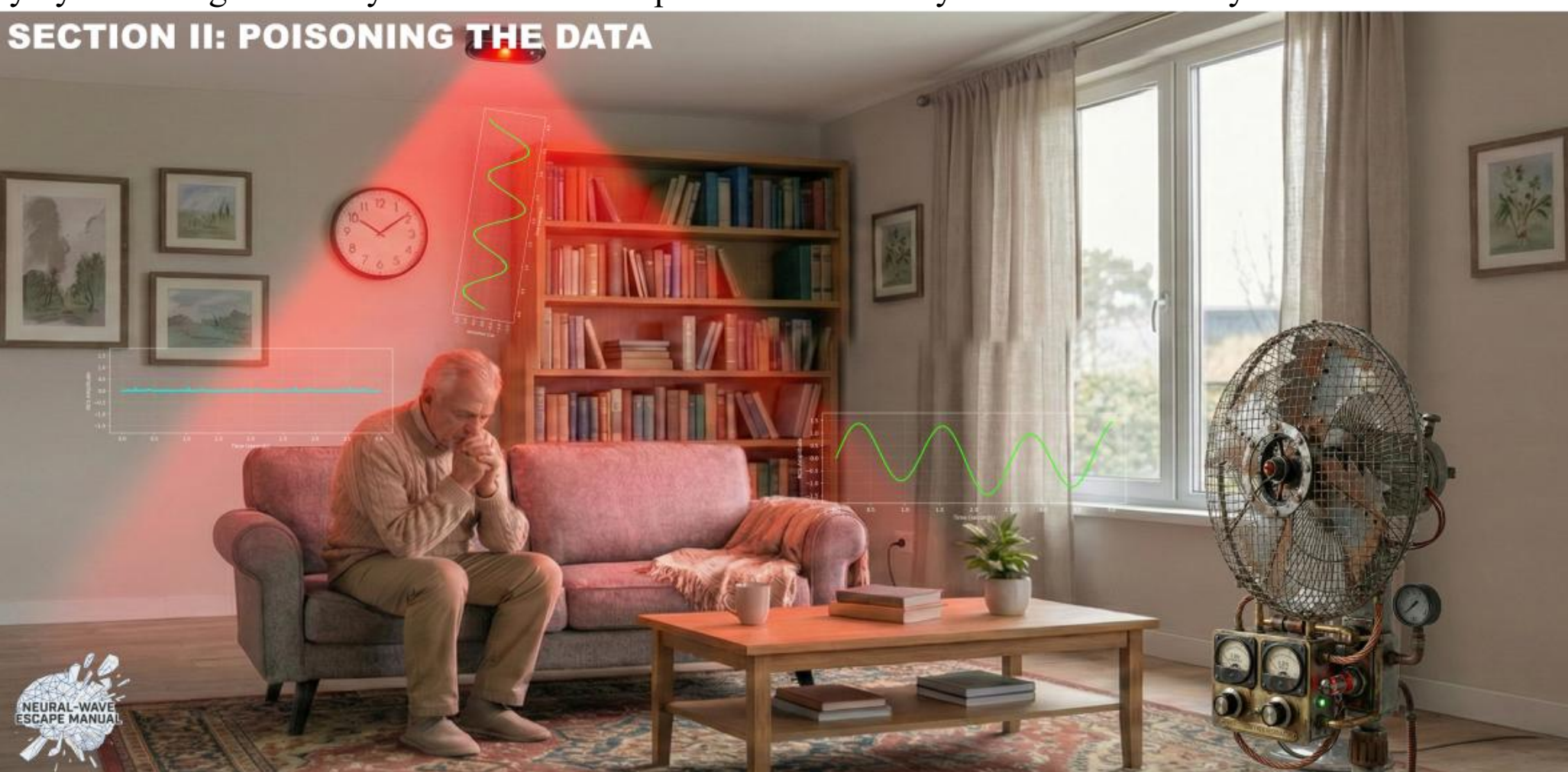

Figure 4. Tactic 02: The Phantom. A modified oscillator mimics a "Hyper-Normal" sinus rhythm. The sensor embraces the perfect signal of the machine, rendering the chaotic, aging body in the corner invisible.

**Safe Zone Assembly — Engineering a Sub-place for Dignity**

Refuse abandons the effort to appease the algorithm, constructing a spatial dead zone where inference cannot penetrate. The manual instructs residents to scavenge conductive refuse—aluminum foil, Mylar, and screens—to retrofit a domestic corner into a makeshift Faraday cage. Designed to attenuate the 60GHz band, this enclosure renders the user as "NULL"—a void in the data stream. Inside this shimmering insulation, the body is finally allowed to stop performing. One can slouch, cry, or curse without these biological states being converted into operational data. Dignity is recovered not by fixing the system, but by walling it out.

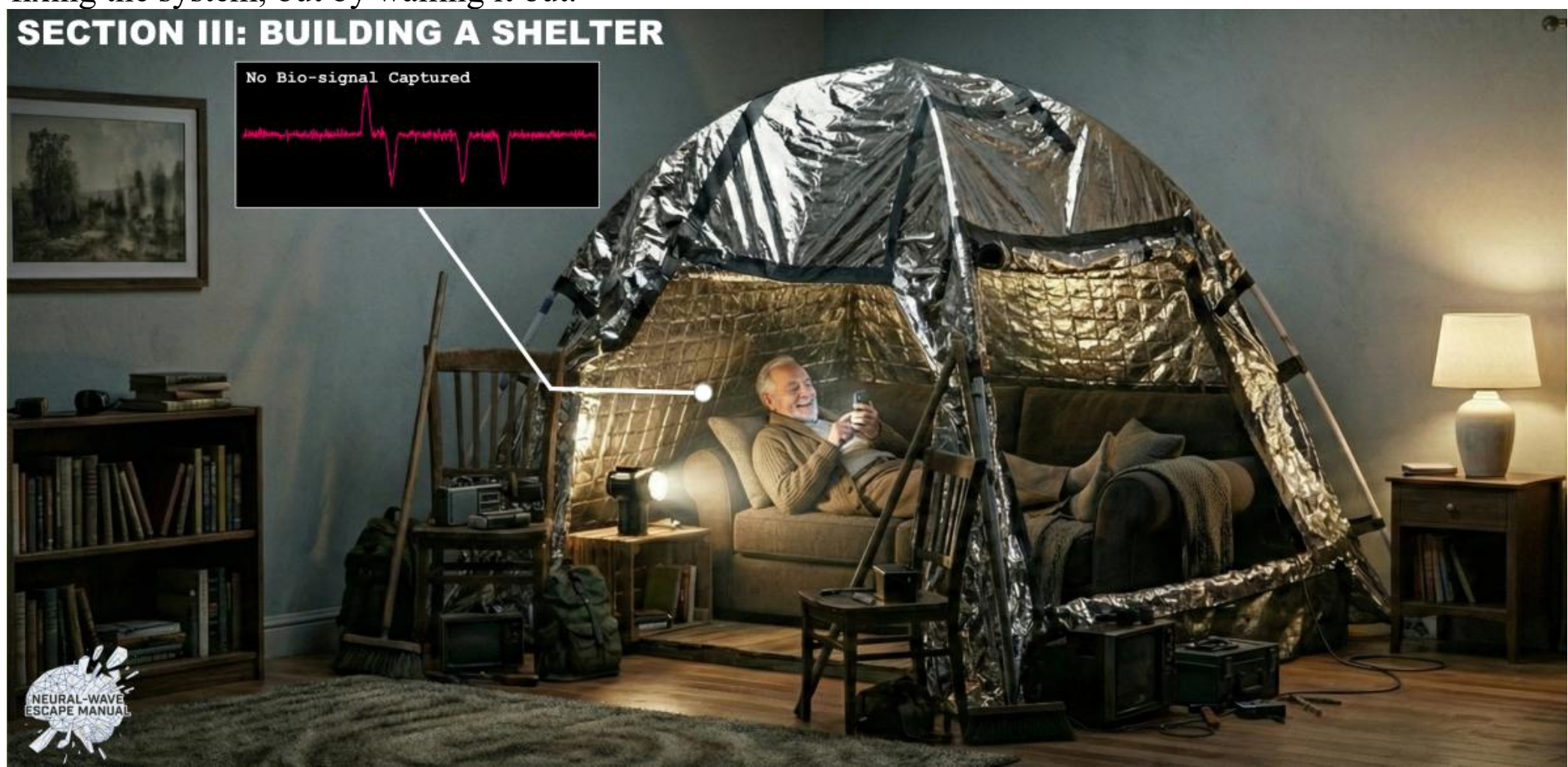

Figure 5. Tactic 03: The Null Space. Aluminum shielding creates a "data void" in the millimeter-wave field. Inside the fortress, the subject reclaims autonomy by becoming spectrally nonexistent.

## 3 DISCUSSION: THE RIGHT TO REMAIN "NOISY"

Returning to Arthur, trapped in his apartment while performing breathing exercises for his ceiling, we see a frictionless future that has become a cage. By placing the Neural-Wave and its Escape Manual in tension, we surface three critical implications for HCI regarding the structural conflicts between machine intelligence and human agency.

First, we challenge the "Accuracy Trap." HCI often assumes that improving inference fidelity resolves ethical tensions. However, Arthur's scenario argues that the deprivation of autonomy is not a failure of accuracy, but a product of it. Even high-fidelity systems fundamentally displace the locus of decision-making. When a system is empowered to act on its inferences (e.g., locking a door), the system's objective data is treated as more "true" than the person's lived reality. The danger is not merely that the machine might be wrong, but that it is unilaterally empowered to be "right," forcing subjects to plead their case against an unappealable black-box score.

Second, we show how ‘passive’ sensing becomes a demand to be legible. While mmWave is often characterized as "unobtrusive," governance-oriented sensing creates a feedback loop compelling active performance. We term this Performative Wellness: a form of coerced labor where the resident must constantly calibrate their body to satisfy the model’s definition of "normality." This shifts the burden of care from the caregiver to the cared-for, who must now work to remain "legible" to the machine [1]. Thus, "passive" sensors actively discipline bodies, enforcing a standardized performance of health that suppresses the messy vitality of actual life.

Finally, we propose Adversarial Literacy as a necessary Digital Right. While cybersecurity frames "adversarial examples" as vulnerabilities, in the context of coercive domestic AI, they are tools for survival. The Escape Manual prototypes this literacy by translating complex signal processing concepts into accessible tools for everyday resistance. By teaching residents to "poison" data streams or construct shields, we suggest that the future of privacy may not lie in policy, but in physics. Building on DiSalvo's account of design as political friction [6], we call for HCI research to explore "Adversarial Design for Non-Experts": tools that help ordinary people obfuscate, interrupt, or selectively blind the sensors around them. The point is not to reject care, but to ensure that care does not erase the right to remain "noisy," opaque, and sovereign.

## 4 CONCLUSION: DON'T FIX THE CODE, BREAK THE SENSOR

This inquiry began with a question: does the pursuit of ever-higher fidelity in AIoT "care" systems systematically trade away the human autonomy to refuse? Our design fiction suggests the answer is yes. By rendering the home completely legible to algorithmic inference, systems like Neural-Wave do not merely monitor life; they begin to curate it, enforcing a standardized performance of wellness that mistakes compliance for health.

The Neural-Wave Quick Escape Manual is more than satirical commentary; it is a material argument for reintroducing friction into AIoT. It demonstrates that when governance becomes infrastructural, resistance must become technical. We are not advocating for an anti-technology future, but rather for a future where domestic technology respects the "right to be left alone" not just as a policy setting, but as a physical reality. As HCI continues to shape the future of aging, we must move beyond designing systems that are merely accurate or "privacy-preserving." We must design systems that are hospitable to the messy, irregular, and sometimes ungovernable vitality of actual human life. This requires a shift from designing for compliance to designing with adversariality—acknowledging that the capacity to obfuscate, confuse, and strategically blind the sensors that watch us is a fundamental prerequisite for dignity in an age of ubiquitous sensing. The most compassionate smart home of the future may well be the one that knows when to look away.

## REFERENCES

[1] Virginia Eubanks. 2018. Automating Inequality: How High-Tech Tools Profile, Police, and Punish the Poor. St. Martin's Press, New York, NY, USA. ISBN: 9781250074317

[2] Anna Brown, Alexandra Chouldechova, Emily Putnam-Hornstein, Andrew Tobin, and Rhema Vaithianathan. 2019. Toward Algorithmic Accountability in Public Services: A Qualitative Study of Affected Community Perspectives on Algorithmic Decision-Making in Child Welfare Services. In Proceedings of the 2019 CHI Conference on Human Factors in Computing Systems (CHI '19). Association for Computing Machinery, New York, NY, USA, Article 41, 12 pages. DOI: 10.1145/3290605.3300271

[3] Anthony Dunne and Fiona Raby. 2013. Speculative Everything: Design, Fiction, and Social Dreaming. MIT Press, Cambridge, MA, USA. ISBN: 9780262019842

[4] James C. Scott. 1985. Weapons of the Weak: Everyday Forms of Peasant Resistance. Yale University Press, New Haven, CT, USA. ISBN: 9780300036411

[5] Finn Brunton and Helen Nissenbaum. 2015. Obfuscation: A User's Guide for Privacy and Protest. MIT Press, Cambridge, MA, USA. DOI: 10.7551/mitpress/9780262029735.001.0001

[6] Carl DiSalvo. 2012. Adversarial Design. MIT Press, Cambridge, MA, USA. DOI: 10.7551/mitpress/8732.001.0001

### Acknowledgements

Images in Figure 1, 3, 4, and 5 were augmented using generative AI tools (Nano Banana & Midjourney) to visualize the design fiction scenarios described in this paper.